\documentclass[prb, showpacs, twocolumn]{revtex4}

\usepackage{graphicx}
\usepackage{dcolumn}
\usepackage{amsmath}
\usepackage{hyperref}

\begin{document}

\title{Aging in the Relaxor Ferroelectric PMN/PT}
\author{Lambert K. Chao}
\author{Eugene V. Colla}
\author{M. B. Weissman}
	\email{mbw@uiuc.edu}
\affiliation{Department of Physics, University of Illinois at
Urbana-Champaign, 1110 West Green Street, Urbana, IL 61801-3080
}

\date{\today}

\begin{abstract}

The relaxor ferroelectric
(PbMn$_{1/3}$Nb$_{2/3}$O$_3$)$_{1-x}$(PbTiO$_3$)$_{x}$, $x=0.1$,
(PMN/PT(90/10)) is found to exhibit several regimes of complicated aging
behavior. Just below the susceptibility peak there is a regime
exhibiting rejuvenation but little memory. At lower temperature, there
is a regime with mainly cumulative aging, expected for simple
domain-growth. At still lower temperature, there is a regime with both
rejuvenation and memory, reminiscent of spin glasses. PMN/PT~(88/12) is
also found to exhibit some of these aging regimes. This qualitative
aging behavior is reminiscent of that seen in reentrant ferromagnets,
which exhibit a crossover from a domain-growth ferromagnetic regime into
a reentrant spin glass regime at lower temperatures. These striking
parallels suggest a picture of competition in PMN/PT~(90/10) between
ferroelectric correlations formed in the domain-growth regime with
glassy correlations formed in the spin glass regime. PMN/PT~(90/10) is
also found to exhibit frequency-aging time scaling of the time-dependent
part of the out-of-phase susceptibility for temperatures 260~K and
below. The stability of aging effects to thermal cycles and field
perturbations is also reported.

\end{abstract}


\maketitle


\section{Introduction}
\label{sec:Intro}

Relaxor ferroelectrics\cite{Smolenskii:59,Cross:87,Cross:94} encompass a
diverse collection of materials. They have crystalline structure and
show medium-range order at high temperature in the form of polar
nanodomains,\cite{Randall:90:JMatRes} but do not enter a ferroelectric
state upon cooling. Instead there is a cooperative faster-than-Arrhenius
kinetic freezing into a glassy relaxor
state.\cite{Glazounov:98:APL,Bokov:99:JPhysCondMatt,Viehland:90:JAP} The
physics underlying this glassy freezing is not well understood, and in
fact may differ amongst the relaxors, as diverse low-temperature aging
behavior in several cubic perovskite relaxors\cite{Colla(Chao:mbw):01}
and uniaxial tungsten-bronze relaxors \cite{Chao(mbw):05} have shown. 

Generically, disordered materials age as they approach thermal
equilibrium and settle into lower free energy states. This aging is
usually exhibited as a decrease in the complex susceptibility $\chi
(\omega, T)=\chi '(\omega, T) - i \chi '' (\omega, T)$ when the sample
is held at fixed temperature $T$ and field ($E$ for dielectrics).
However, while diverse disordered materials may show similar aging under
simple conditions, detailed behavior under complicated aging histories
can shed light on the underlying nature of the disordered states. In the
simplest case, growth of domains give rise to cumulative aging: any
reduction in $\chi (T)$ remains as long as $T$ is kept below a melting
temperature. Here domain size serves as a simple order parameter. In
other cases, aging could involve more complicated order parameters, such
as in spin glasses\cite{Hammann(Vincent):00a} where aging shows
nontrivial effects. Aging a spin glass at a temperature $T_A$ reduces
$\chi (T)$ only in the immediate vicinity of $T_A$ (e.g.,
$|T-T_A|<0.1T_A$), creating an aging ``hole''. So long as $T$ is kept
below $T_A$, there will be memory of this hole (i.e., $\chi (T)$ will
remain reduced in the vicinity of $T_A$). In fact, it is possible to age
at a sequence of decreasing $T_A$'s and measure a series of aging holes
on reheating.\cite{Hammann(Vincent):00a} In addition, spin glasses also
approximately exhibit $\omega t_W$-scaling, where aging of $\chi
(\omega)$ is a function of the product $\omega t_W$ and not $\omega$ and
waiting time $t_W$ separately. This can indicate that the aging and the
response come from the same sort of degrees of freedom, as in
hierarchical kinetic schemes.\cite{Bouchaud:95:JPhysI}

We have previously reported on aging in several cubic perovskite
relaxors of which PbMn$_{1/3}$Nb$_{2/3}$O$_3$ (PMN) and
Pb$_{1-x}$La$_{x}$Zr$_{1-y}$Ti$_{y}$O$_{3}$ (PLZT) showed
spin-glass-like aging at low temperatures below the susceptibility peak.
That behavior contrasted with aging in the uniaxial relaxor
Sr$_x$Ba$_{1-x}$Nb$_2$O$_6$ \cite{Chao(mbw):05} and also with aging in
PMN/PT~(72/28), which develops macroscopic ferroelectric domains. Here
we report on similar aging experiments on PMN/PT~(90/10) and
PMN/PT~(88/12). We find several regimes of zero-field aging behavior in
PMN/PT~(90/10), including a regime of cumulative aging for temperatures
near the susceptibility peak, a regime just below the peak showing
hole-like aging with weak memory effects similar to that seen in some
domain-growth regimes in reentrant
ferromagnets,\cite{Dupuis(Vincent:Hammann):02,%
Vincent:(Hammann:Bouchaud):00} and a second cumulative regime below that
which gradually crosses over into a regime with spin-glass-like behavior
similar to that seen in other cubic perovskites. PMN/PT~(88/12) shows
corresponding regimes except near $T_P$ where aging is hole-like and has
strong memory effects. We also find in PMN/PT~(90/10) approximate
$\omega t_W$-scaling for the time-dependent part of the out-of-phase
susceptibility ($\chi'' (\omega, t)$ with its frequency-dependent
infinite-time asymptotic aging value subtracted) for temperatures
starting near the susceptibility peak all the way down into the
spin-glass-like aging regime. Results from experiments studying the
stability of aging holes to thermal cycles and field perturbations are
also reported. Our interpretation of these results will rely on the
strong parallels between aging in PMN/PT~(90/10) and the reentrant
ferromagnet systems CdCr$_{2x}$In$_{2-2x}$S$_4$ ($x >
0.85$),\cite{Vincent:(Hammann:Bouchaud):00, Dupuis(Vincent:Hammann):02}
which also exhibit competition between ferromagnetic order formed in the
domain growth regime and glassy order formed in the spin glass regime.


\section{Experimental Methods and Results}
\label{sec:ExpMethResults}


\subsection{Techniques}
\label{subsec:Techniques}

The single crystal PMN/PT (90/10) sample under study was grown by the
spontaneous crystallization method at the Institute of Physics, Rostov
State University (Rostov-on-Don, Russia). It was configured as a
capacitor with a thickness of 0.35~mm and electrodes with an area of
$\approx$~5~mm$^2$ on faces perpendicular to the (100) direction. The
single crystal PMN/PT~(88/12) was grown by a modified Bridgeman method
by TRS Technologies, Inc. (State College, PA). It was configured with
electrodes of area of $\approx$~3.5~mm$^2$ on surfaces perpendicular to
the (111) direction, and had a thickness of 5~mm.

Susceptibility was measured with conventional frequency domain
techniques using a current-to-voltage converter built around a
high-impedance op-amp with low input bias current (Analog Devices
AD549LH or Burr-Brown OPA111BM) and a lock-in amplifier. Typically, AC
fields of $\approx$~3~V/cm~rms were applied (within the linear response
regime). Temperature was maintained with a continuous-flow cryostat
capable of temperature sweep rates of up to 20~K/min.


\subsection{Basic Susceptibility and Aging Measurements}
\label{subsec:BasicAging}

\begin{figure}
\includegraphics[width=0.45\textwidth,clip]{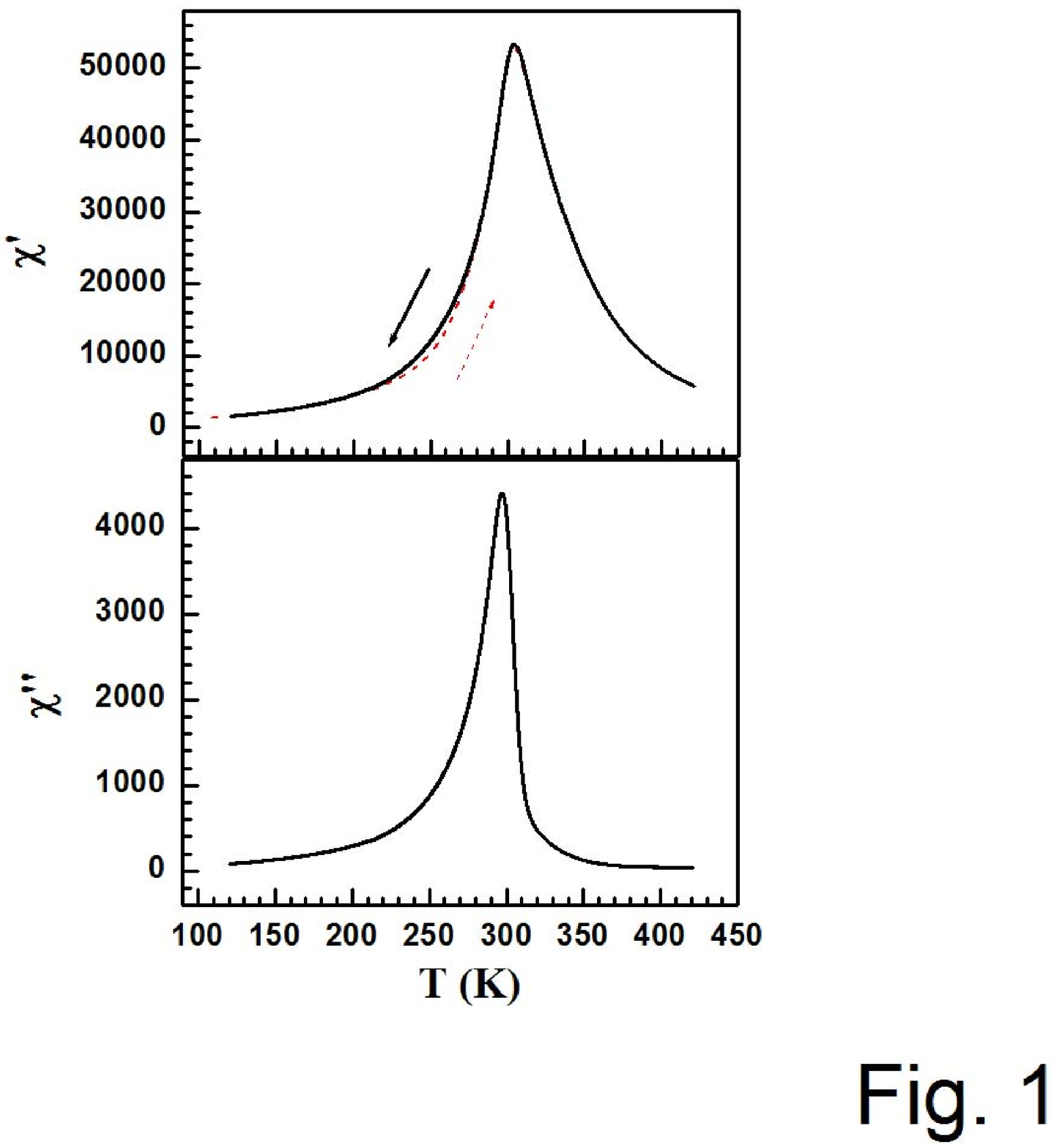}
\caption{
(Color Online) The complex dielectric susceptibility $\chi$ (measured
with an ac field of 3~V/cm~rms at 50~Hz and a $T$-sweep rate of 3~K/min)
is plotted against temperature for PMN/PT~(90/10). Cooling and heating
curves are shown for the in-phase susceptibility.
}\label{fig1}
\end{figure}

The complex dielectric susceptibility $\chi (\omega, T)$
(Fig.~\ref{fig1}) taken at cooling rates of about 3~K/min shows typical
relaxor behavior. We define $T_P=297$~K to be the $\chi'' (T)$ peak of
PMN/PT~(90/10) at 50~Hz. Because of aging, susceptibility curves for
relaxors are somewhat history dependent, and especially so in
PMN/PT~(90/10) because some of the aging is cumulative.


\begin{figure*}
\includegraphics[width=0.75\textwidth,clip]{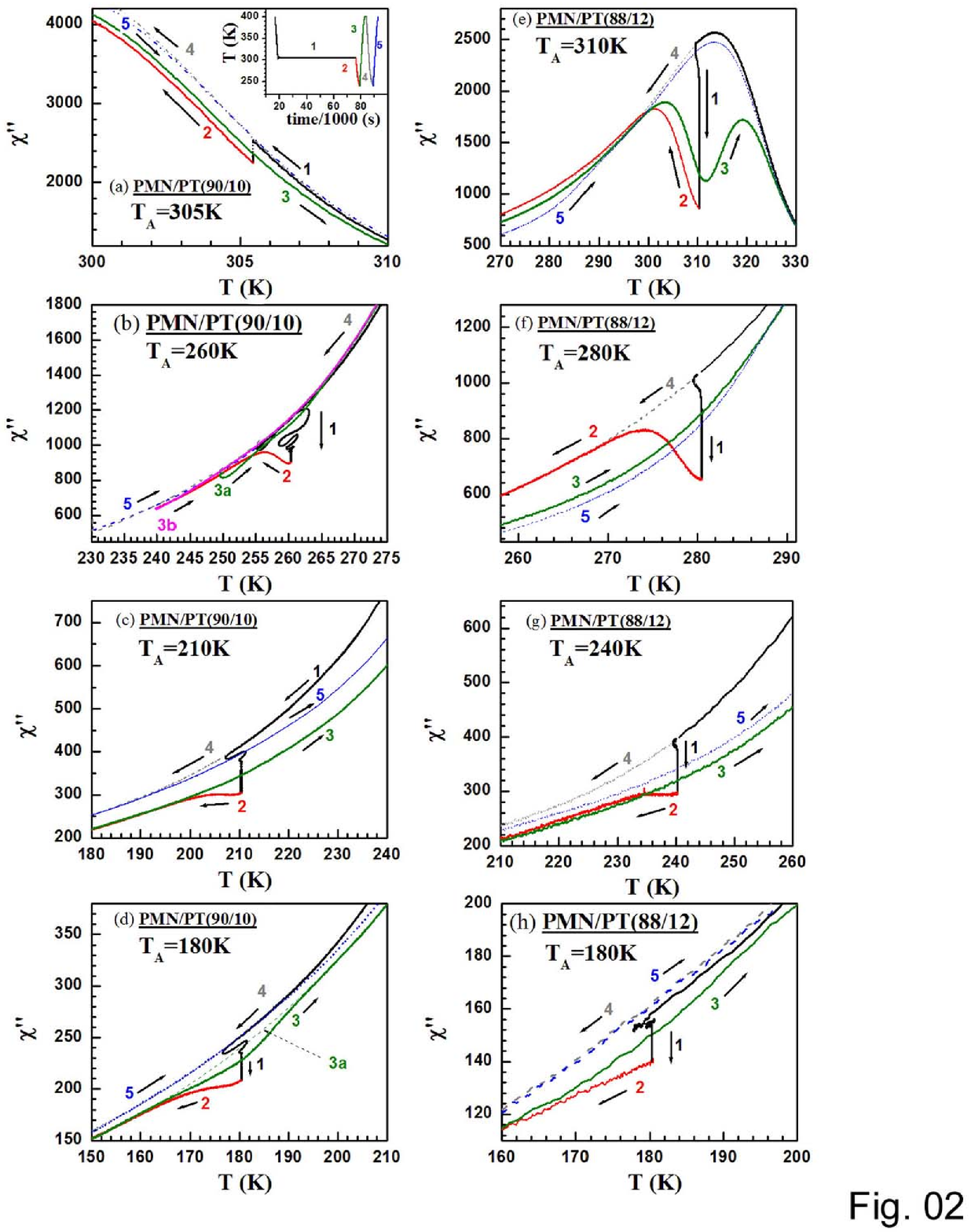}
\caption{
(Color online) Aging behavior of $\chi''$ is shown for several $T_A$.
Inset of panel (a) shows the temperature profile for a typical aging
experiment. Curve 1 is taken on cooling from initial annealing at
temperature $T_{AN}$ to aging temperature $T_A$ held there for aging
time $t_A$. Curve 2 is taken on subsequent cooling to $T_{EX}$ with
immediate reheating back to the $T_{AN}$ along curve 3. Curves 4 and 5
are reference cooling and heating curves (respectively). Measurements
were taken on PMN/PT~(90/10) (PMN/PT~(88/12)) with an ac field of
2.9~V/cm~rms at 50~Hz (1V/cm at 1~kHz) with $T$-sweep rates of about
3~K/min (2~K/min, except (h) which was taken at about 4~K/min). For
PMN/PT~(90/10), $T_{AN}=$ (a,b) 400~K, (c) 340~K, (d) 330~K; $t_A=$
(a,c,d) 16~h, (b) 4~h; $T_{EX}=$ (a) 240~K, (b) 250~K (Curve 3a), 240~K
(Curve 3b), (c,d) 130~K. For PMN/PT~(88/12) $T_{AN}=$~450~K,
$t_A=$~10~h, and $T_{EX}=$ (e) 250~K, (f) 220~K, (g) 180~K, (h) 130~K.
For clarity, damped oscillations from 302-308~K lasting about 0.5~h
arising from the initial temperature overshoot have not been shown in
(a). (b) shows aging curves for two sets of experiments with different
$T_{EX}$'s. Curves 1 and 3 overlay over each other for the two
experiments and only the heating curve 3a of the first experiment is
shown for clarity. Curve 3a in (d) is a polynomial fit of curve 3 with
the ``hole'' from 165-200~K excised and is used to separate the aging
memory into cumulative and hole-like components (see text). The sample
spent about 2~h around 140~K on cooling to $T_{EX}$ in (h) because of a
slowing cooling rate, but with little evidence of a resulting aging
hole. High frequency noise in the aging curve has been removed by
adjacent averaging for (h).
}\label{fig2}
\end{figure*}

In a typical aging memory experiment (see
Fig.~\ref{fig2}(a) inset for temperature history profile),
the sample is cooled to $T_A$ and held there while $\chi (T)$ steadily
decreases. After aging, the sample is thermally cycled to a lower
excursion temperature $T_{EX}$ and then immediately reheated past $T_A$.
For comparison, reference curves are also taken at the same sweep rates
without stopping at $T_A$.


\begin{figure}
\includegraphics[width=0.45\textwidth,clip]{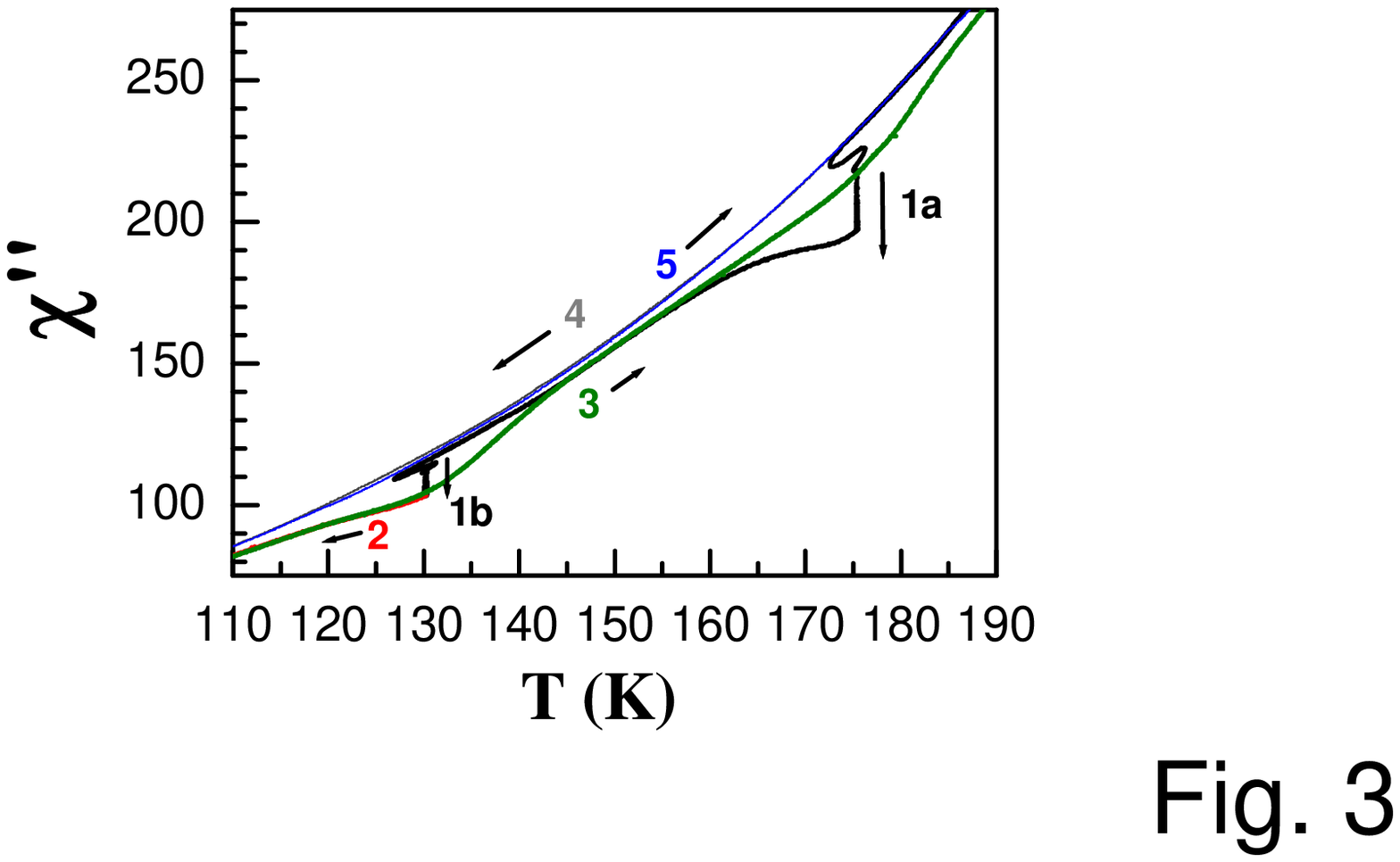}
\caption{
(Color online) The ability to create successive aging holes in $\chi''$
with independent memory is shown. PMN/PT~(90/10) is aged at 175~K (1a)
and subsequently 130~K (1b) for 8~h each, then cooled to 100~K (2) and
immediately reheated (3). $T$-sweep rates were about 3~K/min and
$\chi''$ was measured with an ac field of 2.9~V/cm rms at 50~Hz.
}\label{fig3}
\end{figure}

Our PMN/PT~(90/10) shows several regimes of aging behavior. Above about
330~K, there is little aging. Starting from about 330~K to 290~K, aging
is cumulative: any reduction in $\chi$ remains during the subsequent
thermal cycle (Fig.~\ref{fig2}(a)). Around 280-250~K
(Fig.~\ref{fig2}(b)), lowering $T$ returns $\chi (T)$ to
the reference cooling curve as if there had been no aging, an effect
known as rejuvenation. In this regime, when cooling excursions are
within approximately 10~K of $T_A$, PMN/PT~(90/10) shows weak memory of
the aging hole (there is small dip around $T_A$ on reheating); but if
cooled further below $T_A$, PMN/PT~(90/10) shows no memory and the
reheating curve coincides with the reference curve. Starting around
240~K, PMN/PT~(90/10) gradually crosses over into another cumulative
regime (Fig.~\ref{fig2}(c)): $\chi '$ does not rejuvenate
at all on subsequent cooling after aging, while $\chi''$ does partially
rejuvenate but does not return completely to the reference cooling
curve. This reduction in $\chi ''$ remains on reheating well past $T_A$,
and the cumulative effect of aging in this regime can be seen in the
hysteresis between cooling and heating $\chi(T)$ curves (taken without
pause for aging at fixed $T$) (Fig.~\ref{fig1}). Below 200~K
(Fig.~\ref{fig2}(d)), there is a gradual cross-over to
spin-glass-like aging. On subsequent reheating from $T_{EX}$, $\chi(T)$
shows both cumulative memory (an offset between the reheating curve and
reference heating curve) and hole-like memory (a dip in $\chi(T)$ in the
vicinity of $T_A$ on reheating, reminiscent of memory in
spin glasses\cite{Hammann(Vincent):00a}). We have used a polynomial fit
of the reheating curve (with the dip around $T_A$ excised) to separate
the cumulative and the hole-like components of the memory.
PMN/PT~(88/12) show corresponding aging regimes
(Fig.~\ref{fig2}(f-h)), except near $T_P$ aging is
hole-like with strong memory effects (Fig.~\ref{fig2}(e)).
As in spin glasses, it is possible for multiple independent aging holes
to coexist in the spin-glass-like regime of PMN/PT~(90/10): after aging
at $T_{A1}$ and $T_{A2}$ (less than $T_{A1}$), there is memory of both
aging holes on reheating (Fig.~\ref{fig3}).


\begin{figure}
\includegraphics[width=0.45\textwidth,clip]{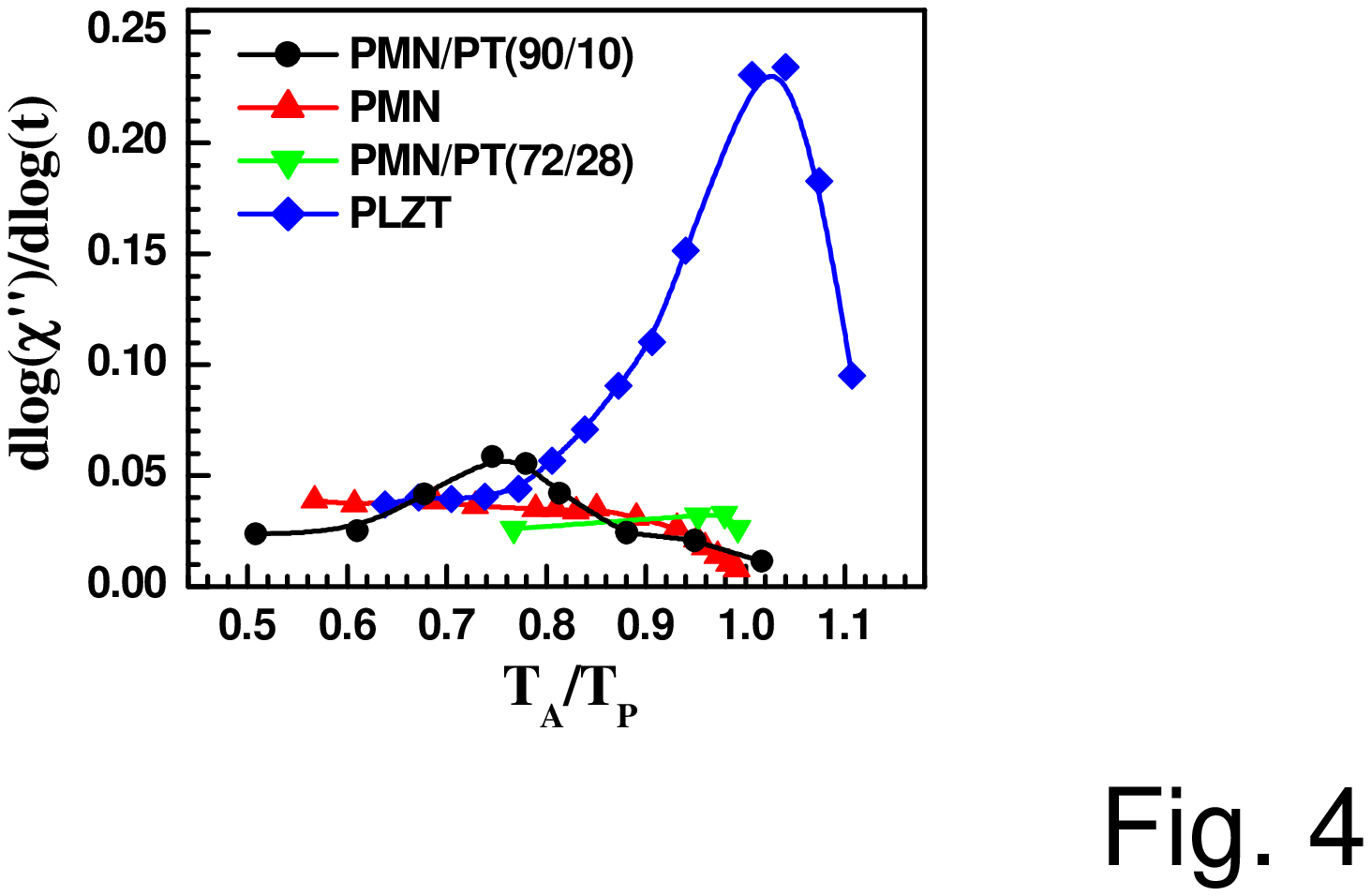}
\caption{
(Color online) Dimensionless aging rate at 10$^4$~s for PMN/PT~(90/10)
(ac frequency 50~Hz), PMN, PMN/PT(72/28) (10~Hz), and PLZT (40~Hz). 
}\label{fig4} 
\end{figure}

In the spin-glass-like aging regime, PMN/PT~(90/10) has aging rates
comparable to other cubic relaxors previously
studied.\cite{Colla(Chao:mbw):01} Figure~\ref{fig4} shows a
dimensionless aging rate $d \ln \chi '' (t_W) / d \ln(t_W)$ at
$t_W=10^4$~s plotted against a normalized aging temperature ($T_A/T_P$)
for PMN/PT (90/10) and several other relaxors. For PMN and PLZT, the
aging rate levels off in the low-$T$ spin-glass-like regime.
PMN/PT~(90/10) shows an increasing aging rate in the cumulative aging
regime that maximizes around 220K, but subsequently falls and levels out
in the spin-glass-like aging regime to rates comparable to the other
cubic relaxors.


\begin{figure*}
\includegraphics[width=0.9\textwidth,clip]{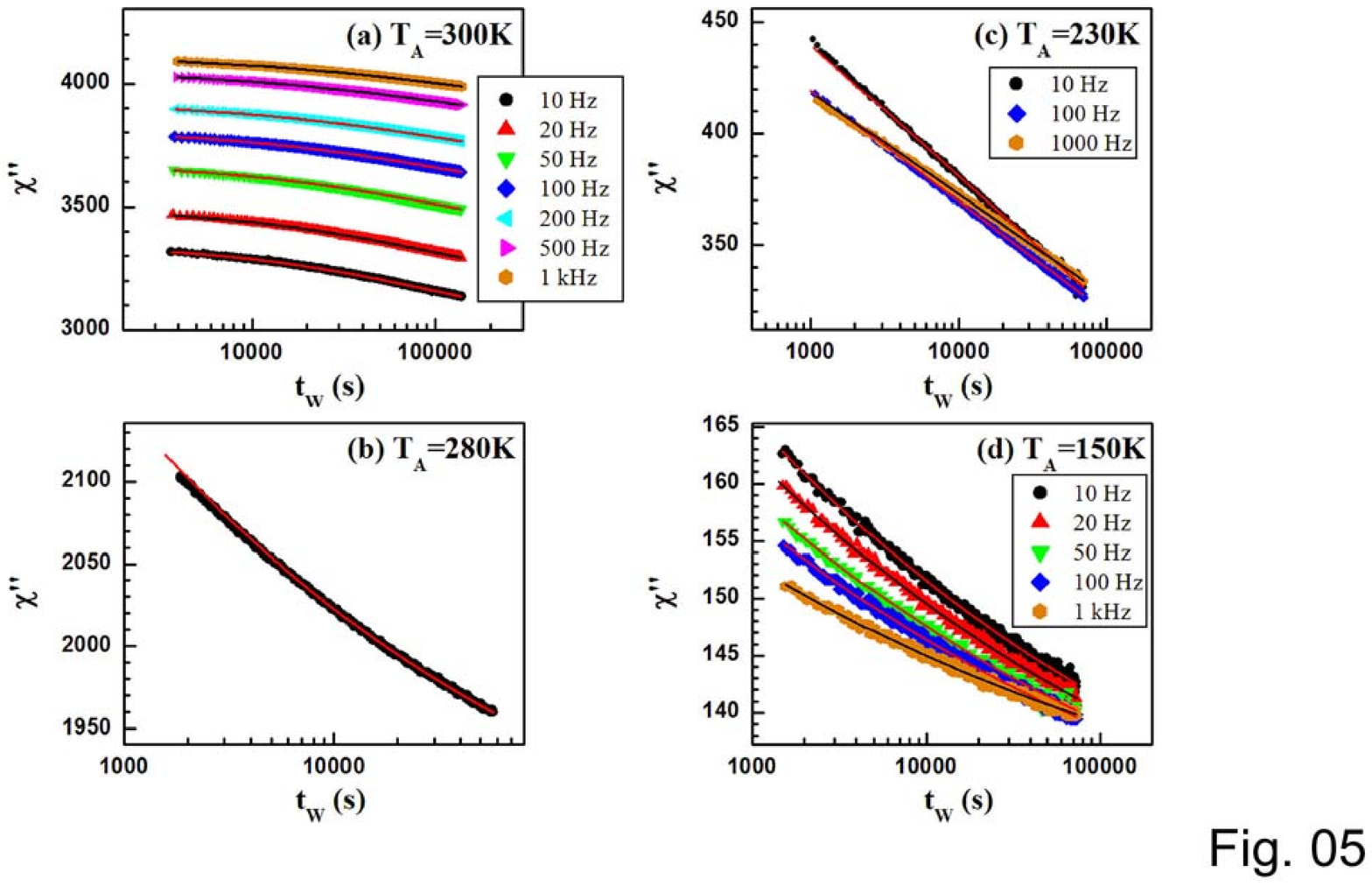}
\caption{
(Color online) Representative $\chi'' (t_W)$ is shown for each of the
aging regimes of PMN/PT~(90/10) along with best fits to various
functional forms (chosen depending on which give physically reasonable
parameters or a smaller chi-squared). (a) $\chi''$ at 300~K was fitted to
a stretched exponential form ($\beta (\omega)$ ranged from 0.63 at 10~Hz
to 0.50 at 1~kHz). (b) 280~K at 50~Hz fitted to power law ($\gamma =
0.19$), (c) 230~K fitted to logarithmic form, and (d) 150~K again to power law
form ($\gamma = 0.14$). (c, d) were measured with the same set of
frequencies as (a), with $\chi(\omega)$ decreasing monotonically as a
function of frequency (only for early $t_W$ in (c)). Data for some
frequencies have been omitted for clarity.
}\label{fig5}
\end{figure*}

Figure~\ref{fig5} shows the decay of $\chi'' (t_W)$ at
several temperatures.  It follows different functional forms depending
on the regime in which aging occurs. Near $T_P$ (330-290~K)
(Fig.~\ref{fig5}(a)), aging fits well to a stretched
exponential form 
\begin{equation} \label{eq:StretchedExpFit}
\chi''(\omega, t_W)=\chi''_o (\omega) \cdot \bigg(1+\alpha (\omega) 
\exp \Big(-\big(\gamma(\omega)t_W\big)^\beta \Big) \bigg),
\end{equation}
while aging in the cumulative regime (240-200~K)
(Fig.~\ref{fig5}(c))fits better to a logarithmic form
\begin{equation} \label{eq:LogFit}
\chi''(t_W)=\chi''_o+g\ln(t_W)
\end{equation}
A power law functional form 
\begin{equation} \label{eq:PwrLawFit}
\chi''(t_W)=\chi''_o+\frac{g}{t_W^\gamma}
\end{equation}
better describes aging in the rejuvenating regime (280-240~K) and aging
well into the spin glass regime (roughly below 150~K), where
domain-growth effects from the cumulative regime have frozen out
(Fig.~\ref{fig5}(b,d), respectively).


\begin{figure}
\includegraphics[width=0.45\textwidth,clip]{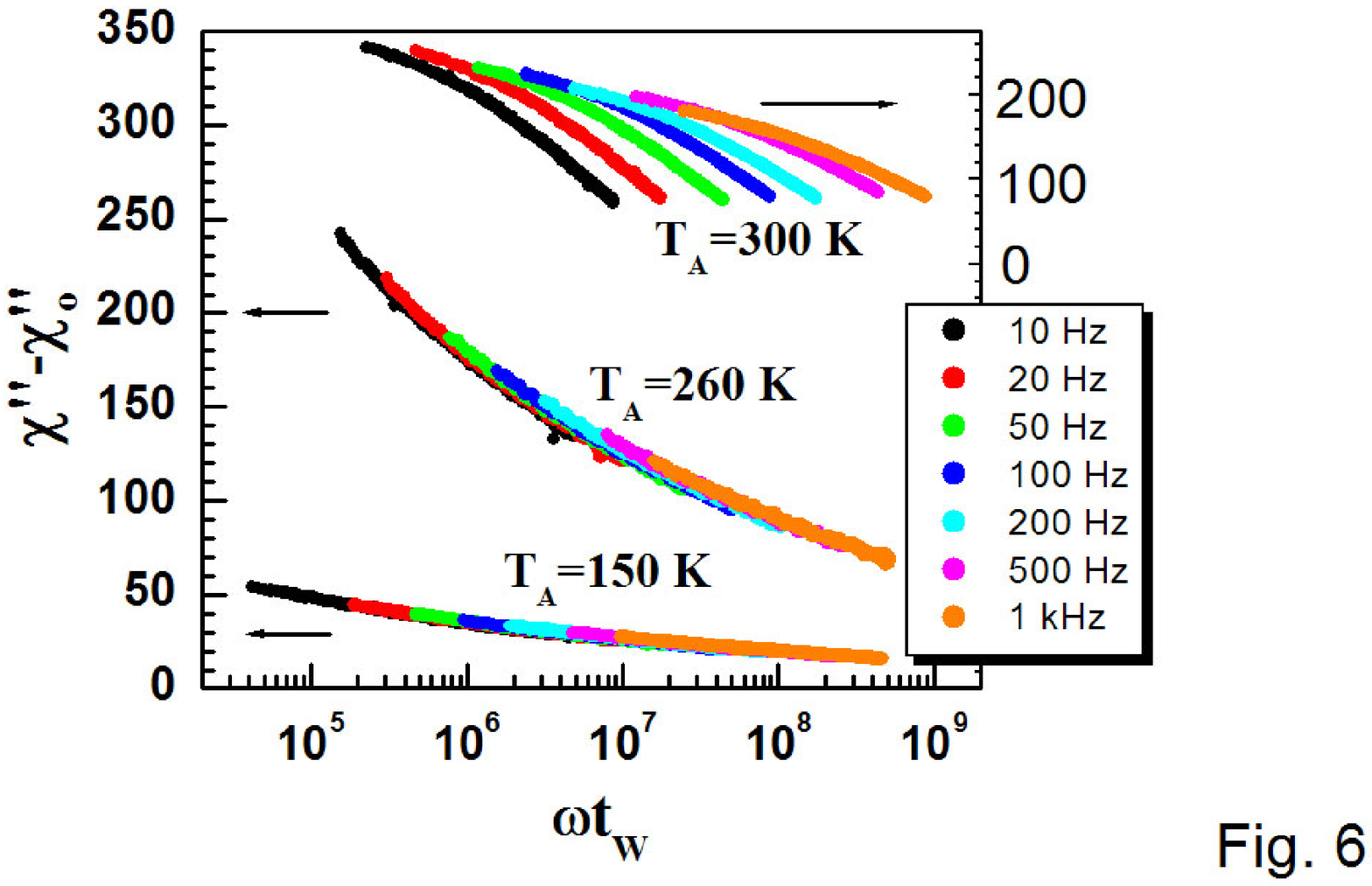}
\caption{
(Color online) $\chi'' (\omega,t_W)$ is plotted against $\omega t_W$ for
several frequencies at 300~K (right axis), 260~K and 150~K (left axis).
Scaling is found in PMN/PT~(90/10) for lower $T$.
}\label{fig6}
\end{figure}

We have previously reported clean $\omega t_W$-scaling of $\chi(\omega,
t_W)$ in PMN and PLZT in the spin-glass-like regime, with very little
frequency dependence of the long-time asymptotic
$\chi''$.\cite{Colla(Chao:mbw):00, Colla(Chao:mbw):01} In
PMN/PT~(90/10), on the other hand, because the asymptotic $\chi''$ is
more frequency dependent, only the time-dependent part of the
susceptibility scales for temperatures up to 260~K (i.e. not just in the
spin-glass-like regime). The aging is well described by 
\begin{equation} \label{eq:wtAgingFit} 
\chi '' (\omega, t_W, T) = \chi''_{o}(\omega, T)
\cdot \left(1+\frac{c(T)}{(\omega t_W)^{\gamma (T)}} \right) 
\end{equation}
Figure~\ref{fig6} shows scaling for $\chi'' (\omega
t_W)-\chi_o'' (\omega)$ for $T = 150$~K in the spin-glass-like regime,
230~K in the lower cumulative regime and 260~K in the domain-growth
regime.  Near $T_P$, where aging follows a stretched exponential decay,
there is no longer scaling.


\subsection{Stability Against Perturbations}
\label{subsec:Stability}

Probing the stability of aging memory against perturbations can provide
information about the detailed nature of the aging and glassy units
involved. To test the stability against temperature perturbations, we
use the aging protocol described above, cycling to a lower temperature
$T_{EX}$ and back after initial aging. The total aging reduction in
$\chi (T_A)$ is compared to the amount of memory of the aging, i.e., the
remaining reduction in $\chi(T_A)$ on reheating after cooling to
$T_{EX}$.

\begin{figure}
\includegraphics[width=0.45\textwidth,clip]{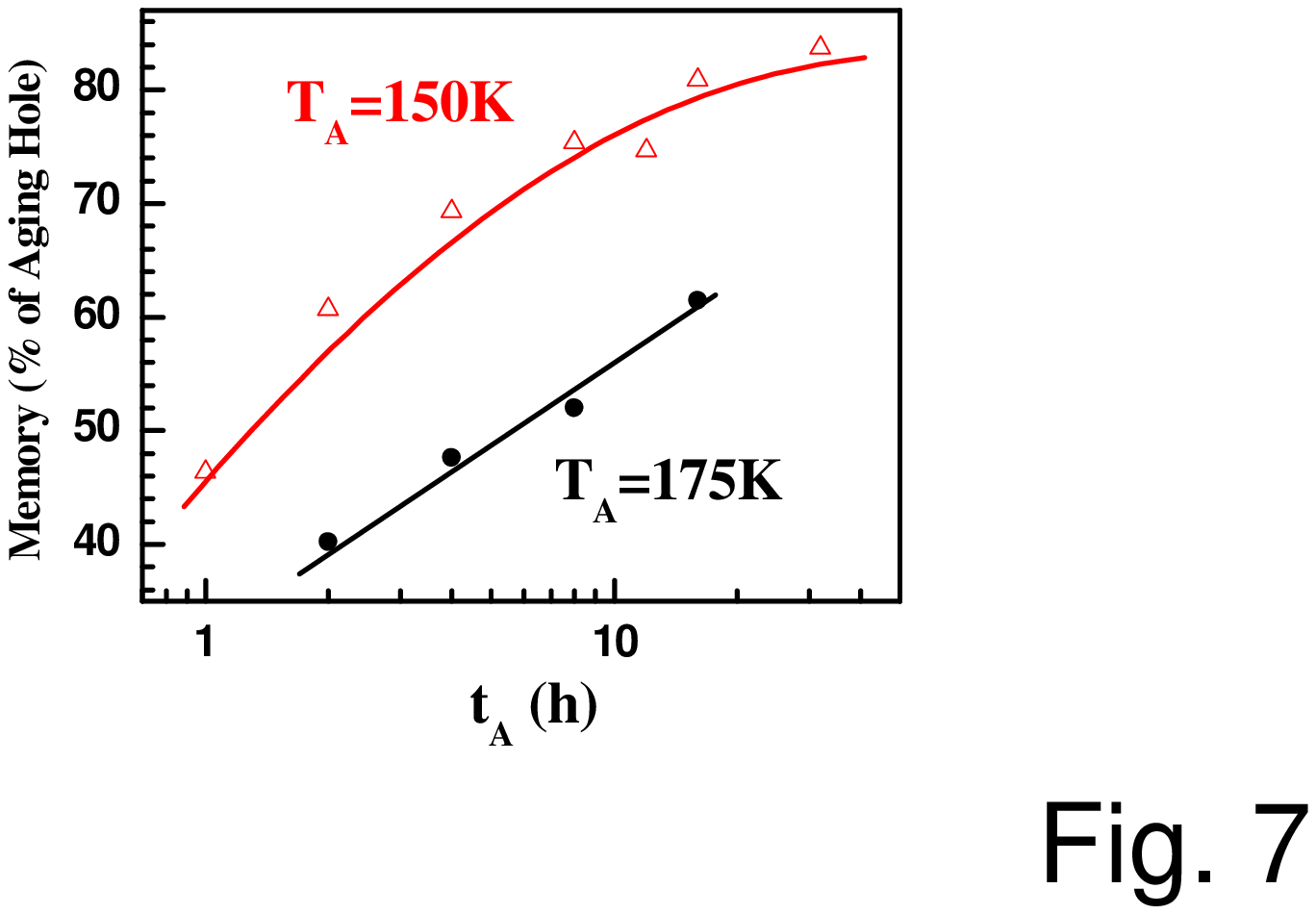}
\caption{
(Color onine) Aging memory is shown as a function of initial aging time
after a cooling cycle to $T_{EX}$ at sweep rates of $\pm$3~K/min.  For
$T_A = $~175~K (150~K), $T_{EX} = $~120~K (105~K). 
}\label{fig7}
\end{figure}

Figure~\ref{fig7} shows the amount of aging memory after
fixed-rate (3~K/min) cycling to $T_{EX}$ as a function of initial aging
time $t_A$ for two aging temperatures. Clearly, any aging
effects are more stable against thermal cycling the longer the initial
aging time.


\begin{figure}
\includegraphics[width=0.45\textwidth,clip]{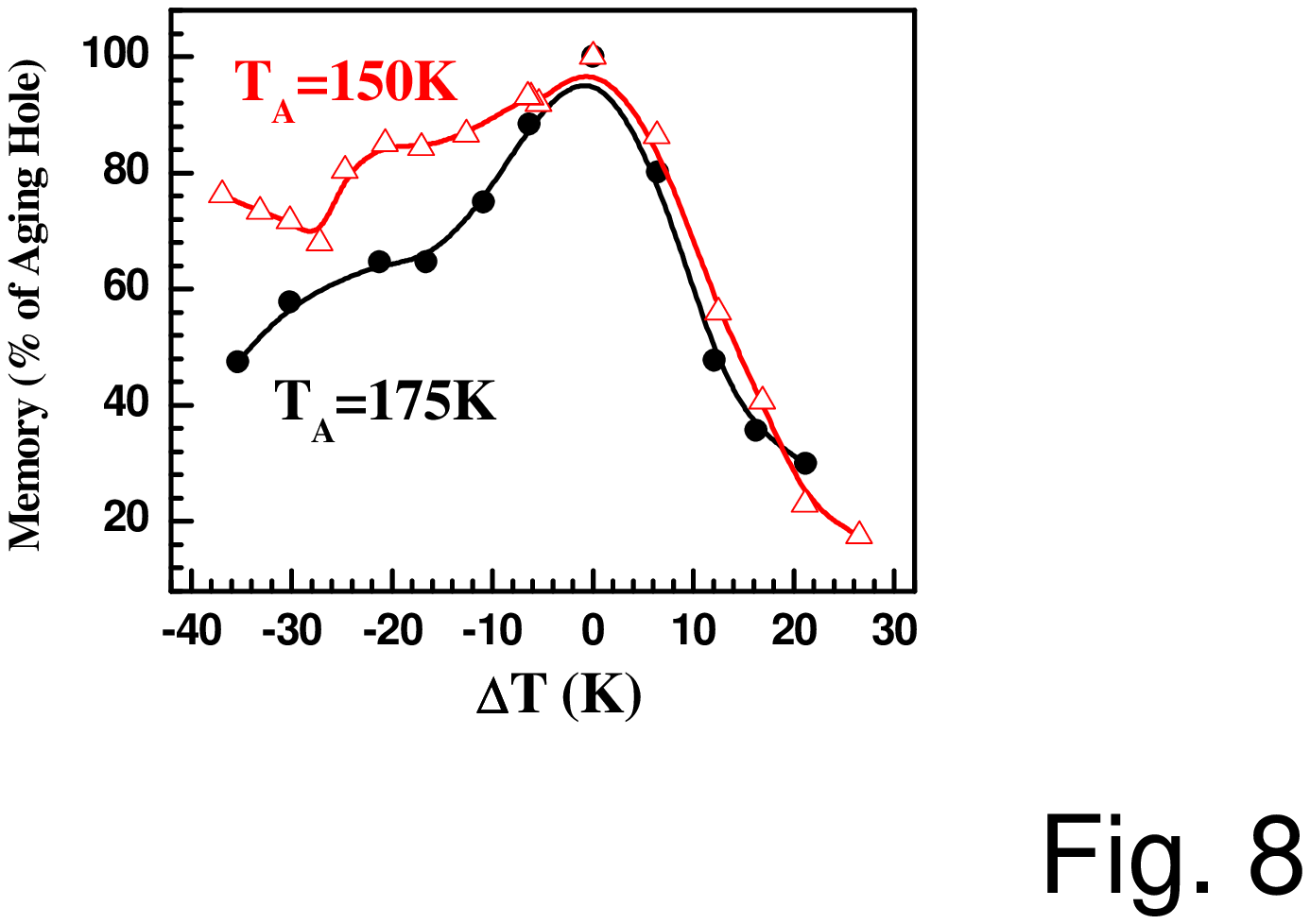}
\caption{
(Color online) Aging memory after small thermal excursions of $\Delta T$
about $T_A$ for PMN/PT~(90/10) at 175~K and 150~K. Sweep rates were
about $\pm$3~K/min.
}\label{fig8}
\end{figure}

Figure~\ref{fig8} shows aging memory after small (positive and negative)
thermal excursions of $\Delta T$ around $T_A$ for two aging temperatures
in the spin glass regime. The behavior is similar to that in the
spin glass regime of other cubic relaxors.\cite{Colla(Chao:mbw):01}
There is a strong asymmetry in the effect of heating and cooling cycles:
heating erases much more memory than cooling. Most memory loss occurs
for small temperature excursions within 10\% of the aging temperature.
While slight cooling erases a large amount of memory, further cooling
does not, leaving some aging memory that is not erased until $T$ is
heated above $T_A$. 


\begin{figure}
\includegraphics[width=0.45\textwidth,clip]{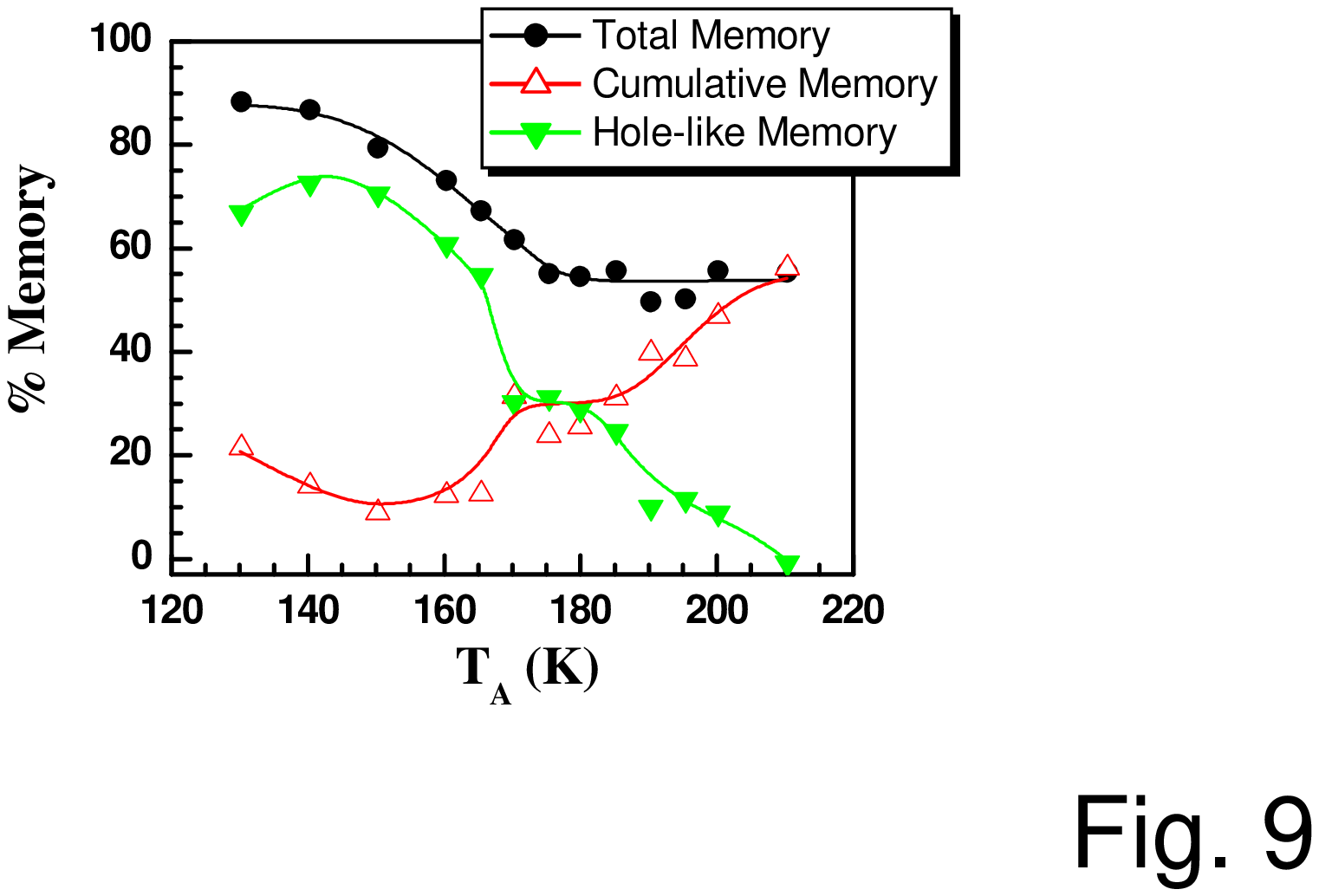}
\caption{
(Color online) Aging memory after cooling cycles to $T_{EX}$ well below
$T_A$ is shown.  Memory has been separated into cumulative and
hole-like components (see Fig.~\ref{fig2}(d)).  For $T_A$
of 175~K and above, $T_{EX}=$~130~K.  For $T_A=$~165~K, $T_{EX}=$~115~K.
$T_{EX}$ is below 95~K for the lowest $T_A$.
}\label{fig9}
\end{figure}

We can probe the temperature dependence of this more persistent memory
using aging experiments with large thermal excursions.
Figure~\ref{fig9} shows memory (in percentage of the
aging reduction remaining) after large thermal excursions as a function
of $T_A$. The sample was aged at $T_A$ for 16~h, then cycled to a fixed
$T_{EX}$ (much lower than any $T_A$) and back. While the absolute
magnitude of the aging decreases with decreasing temperature, the
percentage of that aging retained increases. Since there is a cumulative
as well as a hole-like part to the aging memory in this regime
(Fig.~\ref{fig2}(d)), we have separated the total memory
into its cumulative and hole-like parts using a polynomial fit as
before. The crossover from mostly cumulative to mostly hole-like aging
is evident.


\begin{figure}
\includegraphics[width=0.45\textwidth,clip]{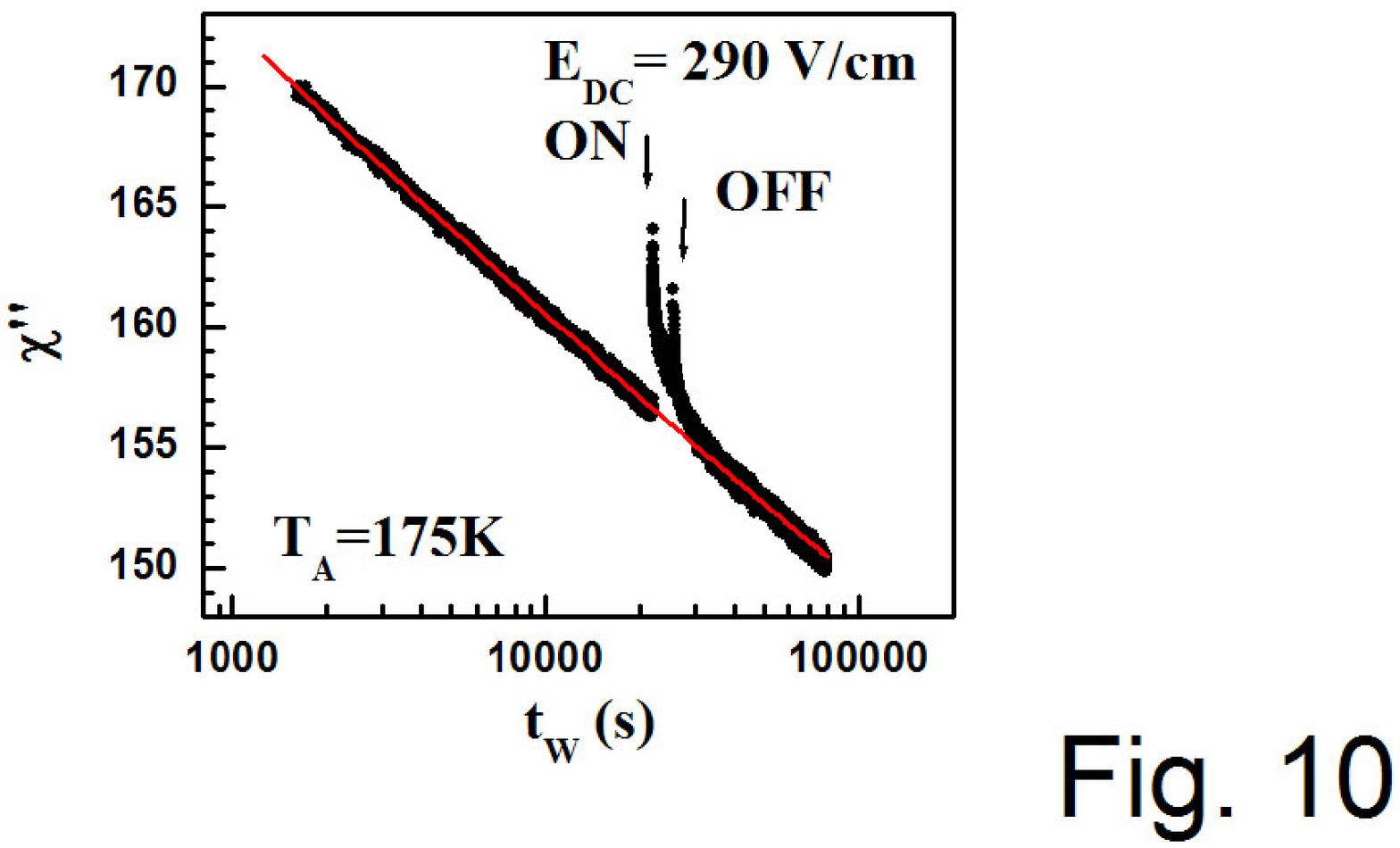}
\caption{
(Color online) Effect of a 1~h field application after aging in zero
field for 6~h on $\chi''$ is shown for PMN/PT~(90/10).
}\label{fig10}
\end{figure}

\begin{figure}
\includegraphics[width=0.45\textwidth,clip]{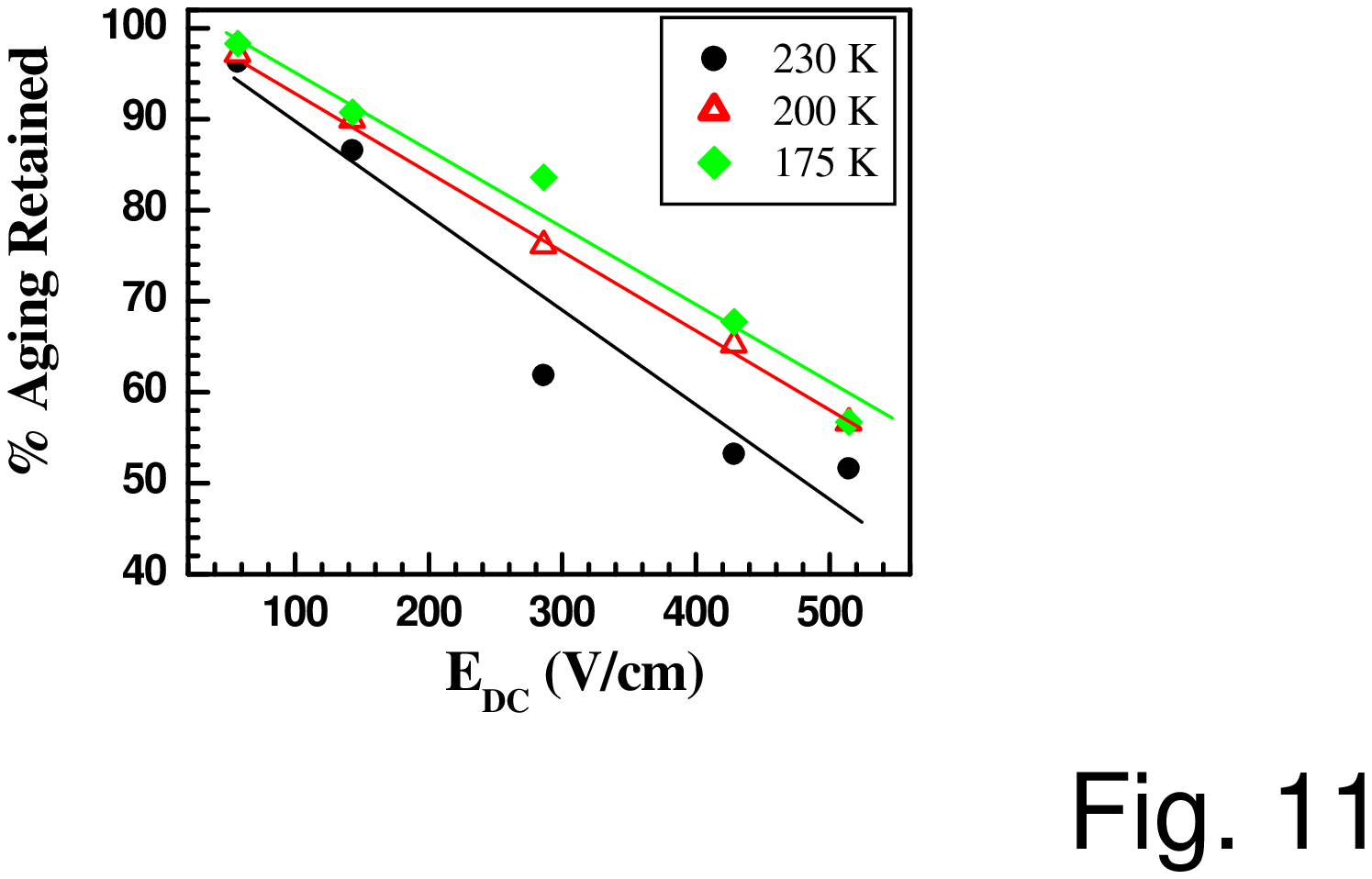}
\caption{
(Color online) A measure of the effect of a sudden application of an
electric field (as in Fig.~\ref{fig10}) on aging of $\chi''$
is shown as a function of applied field. We define the amount of initial
zero-field aging reduction of $\chi''$ remaining 30~s after the DC field
application as a measure of aging recovery. 
}\label{fig11}
\end{figure}

We also studied the effects of field perturbations on aging in the
spin glass regime. This gives an estimate of the field scale necessary
to disrupt any established aging order. In a typical experiment,
PMN/PT~(90/10) is cooled and aged for 3~h in zero field. A DC field is
then applied for 1~h, and then turned off and the sample allowed to
relax in zero field (Fig.~\ref{fig10} at 175~K with a field of
286~V/cm). We see that applying the field erases some of the aging done
in zero field, but after the field is turned off, the sample returns
asymptotically to the initial zero-field aging curve. The loss of memory
on turning on $E$ is a monotonic function of $\Delta E$ reaching 50\% loss
at roughly 500~V/cm (Fig.~\ref{fig11}).


\section{Discussion}
\label{sec:Discussion}

It should be noted that previous work by Kircher \emph{et al.} on a
ceramic PMN/PT~(90/10) sample concentrated near the susceptibility peak
show results qualitatively different from those of our PMN/PT~(90/10)
sample: hole-like aging with stronger memory effects near $T_P$, no
$\omega t_W$-scaling, and a $\chi''$ peak nearly 20~K below ours (280~K
compared to 296~K, measured at 20~Hz). While the difference in $T_P$'s
suggest a substantial difference between the two 10\% samples (one
possibility being ceramic strain effects), it is unclear how they could
explain such qualitatively different aging behavior near $T_P$,
considering that our PMN/PT~(88/12) sample shows hole-like aging similar
to the Kircher sample near $T_P$.

Two of the aging regimes of PMN/PT~(90/10) (the rejuvenating regime just
below $T_P$ showing little memory and the spin-glass-like regime at low
temperatures) directly parallel regimes in reentrant spin glasses, such
as CdCr$_{2x}$In$_{2-2x}$S$_4$ ($x >
0.85$).\cite{Vincent:(Hammann:Bouchaud):00, Dupuis(Vincent:Hammann):02}
For weak disorder, these materials cool from a paramagnetic phase into a
``ferromagnetic'' regime with short range order and then subsequently
into a low-temperature spin glass regime. The ``ferromagnetic'' phase
shows aging like our rejuvenating regime: rejuvenation and weak memory
easily erased by small cooling excursions. Similarly the reentrant
spin glass regime shows typical spin glass behavior, as does ours. As in
PMN/PT~(90/10), both regimes show $\omega t_W$-scaling of the
time-dependent part of the susceptibility. However, aging in the
crossover between these two magnetic regimes were not reported in
detail, so it is unknown whether there is a corresponding distinct
cumulative aging regime in between the two, as seen in our
PMN/PT~(90/10) system and also in polymers\cite{Bellon:00:EPL} and
ferroelectrics.\cite{Mueller:04:EPL, Alberici-Kious:98:PRL,%
Bouchaud:EPJB:01}

It has been suggested in passing that the relaxors may be related to
reentrant spin glasses.\cite{Vakhrushev(Egami):96} Motivated by the
insensitivity (compared to spin glasses) of aging effects to field
perturbations\cite{Colla(Chao:mbw):01} along with a Barkhausen noise
temperature dependence\cite{Colla(Chao:mbw):02} which suggest that the
units responsible for glassy behavior in PMN are much smaller than any
nanodomains, we have proposed a picture of the cubic relaxors
\cite{Weissman:03Williamsburg,Colla(mbw):05} where local ferroelectric
regions freeze in a glass of canted moments, much like a reentrant xy
spin glass.\cite{Ryan(Coey):87, Senoussi:88} Dipole moments orthogonal to
the local polarization observed in PMN \cite{Dkhil:01, Egami:99} are
possible candidates for this glassy freezing, and it is known that
glassy regions exist between nanodomains,\cite{Vakhrushev:97, Blinc:03}
although the extent to which the glassy and the ferro regions overlap
remains unclear. The strength and nature of the random coupling between
the glassy and ferro components would then account for the relaxor
behavior. The orthogonal components found in neutron pair distribution
functions \cite{Jeong:05:PRL} could be the unit-scale analog of the
mesoscopic tweed-like structuring seen in PMN/PT~(65/35) both above and
well below the martensitic
transformation.\cite{Viehland:95:APL,Dai:94:JAP} Viehland \emph{et al.}
has noted\cite{Viehland:95:APL} in particular the relevance of a theory
\cite{Kartha:91:PRL, Sethna:92:PhysScript} mapping the pre-martensitic
tweed Hamiltonian onto a spin glass Hamiltonian.

The similarity in aging behavior between PMN/PT~(90/10) and reentrant
spin glasses is consistent with this idea. We suspect that the underlying
physical picture is directly analogous to that in the reentrant
spin glasses. Just below $T_P$, where the effective ferroelectric
interaction energies between weakly polarized regions are weaker than
the local random fields found in the relaxors, aging involves
short-range ``domains'' or polar regions aligning and forming detailed
patterns under their local random fields. During the subsequent cooling
cycle after aging, the domains grow larger as ferroelectric correlations
grow stronger and the free energy ground state the domains were
equilibrating towards is no longer favored, resulting in rejuvenation.
Any detailed domain patterns formed during aging are wiped out by domain
growth, so there is no memory on reheating. At lower temperatures, the
ferroelectric correlations continue to grow resulting in
domain-growth-like cumulative aging. These large domains then gradually
freeze out on further cooling, leaving behind only glassy degrees of
freedom associated with displacements orthogonal to the domain
polarization, much like the onset of the reentrant spin glass phase.

Other aspects of PMN/PT~(90/10) behavior are also consistent with its
increased ferroelectric content compared to PMN. In pure PMN, $\chi
(\omega)$ showed $\omega t_W$-scaling\cite{Colla(Chao:mbw):01} at low
temperature similar to spin glasses\cite{Hammann(Vincent):00a} and
suggests that the ferro regions responsible for the response are tightly
coupled to glassy order responsible for the aging. Hierarchical schemes
are believed to explain dynamics of spin glasses as well as pinned
domain walls\cite{Balents:96:JPhysI,Dupuis(Vincent:Hammann):02} Here,
for PMN/PT~(90/10) with stronger ferroelectric correlations, it is not
surprising that the coupling between the ferro and glassy components has
changed and $\omega t_W$-scaling is different from that of pure PMN. The
field required to perturb established aging is also comparable to that
of pure PMN,\cite{Colla(Chao:mbw):01} and Barkhausen noise experiments
show a similar temperature dependence and dipole moment step
size.\cite{Chao(mbw):03}

So far we have PMN and PMN/PT~(90/10) showing behavior consistent with
the orthogonal-glass picture, whereas the uniaxial relaxor
Sr$_x$Ba$_{1-x}$Nb$_2$O$_6$,\cite{Chao(mbw):05} with no local orthogonal
moments, does not show a spin-glass-like regime at low temperatures. For
PMN/PT, increasing PT concentration has the effect of tuning the
strength of the ferroelectric correlations from that of pure PMN.
PMN/PT~(90/10) has been seen to show some mottled ``domain patterning''
but no macroscopic ferroelectric domains using TEM\cite{Viehland:95:APL}
and polarized optical imaging.\cite{Viehland:04:JAP} The differences in
behavior between pure PMN and PMN/PT~(90/10) are consistent with a
slight increase in ferroelectric correlations. The aging behavior shows
a rejuvenating regime as well as a domain-growth cumulative regime and
there is $\omega t_W$-scaling of only the dynamic part of the
susceptibility scales ($\chi''-\chi''_o$) (as opposed to scaling of
$\chi''$ in pure PMN). At the far extreme, PMN/PT~(72/28) shows
macroscopic ferroelectric domains \cite{Viehland:04:JAP} and
rejuvenating aging with no memory.\cite{Colla(Chao:mbw):01} Future work
further exploring this idea of interplay between ferro and glassy order
using $x$ in the PMN/PT system to tune the ferro strength will be
interesting. We hope that the aging properties of relevant theoretical
models\cite{Kartha:91:PRL, Sethna:92:PhysScript} can be calculated for
comparison with experimental results.

\section*{Acknowledgement} 

This work was funded by NSF DMR 02-40644 and used facilities of the
Center for Microanalysis of Materials, University of Illinois, which is
partially supported by the U.S. Department of Energy under grant
DEFG02-91-ER4543. We thank the Institute of Physics, Rostov State
University for the PMN/PT~(90/10) sample and D.~Viehland for the
PMN/PT~(88/12) sample.

\bibliography{ChaoBib}

\begin{thebibliography}{36}
\expandafter\ifx\csname natexlab\endcsname\relax\def\natexlab#1{#1}\fi
\expandafter\ifx\csname bibnamefont\endcsname\relax
  \def\bibnamefont#1{#1}\fi
\expandafter\ifx\csname bibfnamefont\endcsname\relax
  \def\bibfnamefont#1{#1}\fi
\expandafter\ifx\csname citenamefont\endcsname\relax
  \def\citenamefont#1{#1}\fi
\expandafter\ifx\csname url\endcsname\relax
  \def\url#1{\texttt{#1}}\fi
\expandafter\ifx\csname urlprefix\endcsname\relax\def\urlprefix{URL }\fi
\providecommand{\bibinfo}[2]{#2}
\providecommand{\eprint}[2][]{\url{#2}}

\bibitem[{\citenamefont{Smolenskii and Agranovskaya}(1959)}]{Smolenskii:59}
\bibinfo{author}{\bibfnamefont{G.~A.} \bibnamefont{Smolenskii}}
  \bibnamefont{and} \bibinfo{author}{\bibfnamefont{A.~I.}
  \bibnamefont{Agranovskaya}}, \bibinfo{journal}{Sov. Phys. Solid State}
  \textbf{\bibinfo{volume}{1}}, \bibinfo{pages}{1429} (\bibinfo{year}{1959}).

\bibitem[{\citenamefont{Cross}(1987)}]{Cross:87}
\bibinfo{author}{\bibfnamefont{L.~E.} \bibnamefont{Cross}},
  \bibinfo{journal}{Ferroelectrics} \textbf{\bibinfo{volume}{76}},
  \bibinfo{pages}{241} (\bibinfo{year}{1987}).

\bibitem[{\citenamefont{Cross}(1994)}]{Cross:94}
\bibinfo{author}{\bibfnamefont{L.~E.} \bibnamefont{Cross}},
  \bibinfo{journal}{Ferroelectrics} \textbf{\bibinfo{volume}{151}},
  \bibinfo{pages}{305} (\bibinfo{year}{1994}).

\bibitem[{\citenamefont{Randall et~al.}(1990)\citenamefont{Randall, Bhalla,
  Shrout, and Cross}}]{Randall:90:JMatRes}
\bibinfo{author}{\bibfnamefont{C.~A.} \bibnamefont{Randall}},
  \bibinfo{author}{\bibfnamefont{A.~S.} \bibnamefont{Bhalla}},
  \bibinfo{author}{\bibfnamefont{T.~R.} \bibnamefont{Shrout}},
  \bibnamefont{and} \bibinfo{author}{\bibfnamefont{L.~E.} \bibnamefont{Cross}},
  \bibinfo{journal}{J. Mater. Res.} \textbf{\bibinfo{volume}{5}},
  \bibinfo{pages}{829} (\bibinfo{year}{1990}).

\bibitem[{\citenamefont{Glazounov and Tagantsev}(1998)}]{Glazounov:98:APL}
\bibinfo{author}{\bibfnamefont{A.~E.} \bibnamefont{Glazounov}}
  \bibnamefont{and} \bibinfo{author}{\bibfnamefont{A.~K.}
  \bibnamefont{Tagantsev}}, \bibinfo{journal}{Appl. Phys. Lett.}
  \textbf{\bibinfo{volume}{73}}, \bibinfo{pages}{856} (\bibinfo{year}{1998}).

\bibitem[{\citenamefont{Bokov et~al.}(1999)\citenamefont{Bokov, Leshchenko,
  Malitskaya, and Raevski}}]{Bokov:99:JPhysCondMatt}
\bibinfo{author}{\bibfnamefont{A.~A.} \bibnamefont{Bokov}},
  \bibinfo{author}{\bibfnamefont{M.~A.} \bibnamefont{Leshchenko}},
  \bibinfo{author}{\bibfnamefont{M.~A.} \bibnamefont{Malitskaya}},
  \bibnamefont{and} \bibinfo{author}{\bibfnamefont{I.~P.}
  \bibnamefont{Raevski}}, \bibinfo{journal}{J. Phys. Condens. Matter}
  \textbf{\bibinfo{volume}{11}}, \bibinfo{pages}{4899} (\bibinfo{year}{1999}).

\bibitem[{\citenamefont{Viehland et~al.}(1990)\citenamefont{Viehland, Jang,
  Cross, and Wuttig}}]{Viehland:90:JAP}
\bibinfo{author}{\bibfnamefont{D.}~\bibnamefont{Viehland}},
  \bibinfo{author}{\bibfnamefont{S.~J.} \bibnamefont{Jang}},
  \bibinfo{author}{\bibfnamefont{L.~E.} \bibnamefont{Cross}}, \bibnamefont{and}
  \bibinfo{author}{\bibfnamefont{M.}~\bibnamefont{Wuttig}},
  \bibinfo{journal}{J. Appl. Phys.} \textbf{\bibinfo{volume}{68}},
  \bibinfo{pages}{2916} (\bibinfo{year}{1990}).

\bibitem[{\citenamefont{Colla et~al.}(2001)\citenamefont{Colla, Chao, and
  Weissman}}]{Colla(Chao:mbw):01}
\bibinfo{author}{\bibfnamefont{E.~V.} \bibnamefont{Colla}},
  \bibinfo{author}{\bibfnamefont{L.~K.} \bibnamefont{Chao}}, \bibnamefont{and}
  \bibinfo{author}{\bibfnamefont{M.~B.} \bibnamefont{Weissman}},
  \bibinfo{journal}{Phys. Rev. B} \textbf{\bibinfo{volume}{63}},
  \bibinfo{pages}{134107} (\bibinfo{year}{2001}).

\bibitem[{\citenamefont{Chao et~al.}(2005)\citenamefont{Chao, Colla, Weissman,
  and Viehland}}]{Chao(mbw):05}
\bibinfo{author}{\bibfnamefont{L.~K.} \bibnamefont{Chao}},
  \bibinfo{author}{\bibfnamefont{E.~V.} \bibnamefont{Colla}},
  \bibinfo{author}{\bibfnamefont{M.~B.} \bibnamefont{Weissman}},
  \bibnamefont{and} \bibinfo{author}{\bibfnamefont{D.~D.}
  \bibnamefont{Viehland}}, \bibinfo{journal}{Phys. Rev. B}
  \textbf{\bibinfo{volume}{72}}, \bibinfo{pages}{134105}
  (\bibinfo{year}{2005}).

\bibitem[{\citenamefont{J.Hammann et~al.}(2000)\citenamefont{J.Hammann,
  E.Vincent, V.Dupuis, M.Alba, M.Ocio, and
  J.-P.Bouchaud}}]{Hammann(Vincent):00a}
\bibinfo{author}{\bibnamefont{J.Hammann}},
  \bibinfo{author}{\bibnamefont{E.Vincent}},
  \bibinfo{author}{\bibnamefont{V.Dupuis}},
  \bibinfo{author}{\bibnamefont{M.Alba}},
  \bibinfo{author}{\bibnamefont{M.Ocio}}, \bibnamefont{and}
  \bibinfo{author}{\bibnamefont{J.-P.Bouchaud}}, \bibinfo{journal}{J. Phys.
  Soc. Jpn., Suppl. A} \textbf{\bibinfo{volume}{69}}, \bibinfo{pages}{206}
  (\bibinfo{year}{2000}), \eprint{cond-mat/9911269}.

\bibitem[{\citenamefont{Bouchaud and Dean}(1995)}]{Bouchaud:95:JPhysI}
\bibinfo{author}{\bibfnamefont{J.~P.} \bibnamefont{Bouchaud}} \bibnamefont{and}
  \bibinfo{author}{\bibfnamefont{D.~S.} \bibnamefont{Dean}},
  \bibinfo{journal}{J. Phys. I France} \textbf{\bibinfo{volume}{5}},
  \bibinfo{pages}{265} (\bibinfo{year}{1995}).

\bibitem[{\citenamefont{Dupuis et~al.}(2002)\citenamefont{Dupuis, Vincent,
  Alba, and Hammann}}]{Dupuis(Vincent:Hammann):02}
\bibinfo{author}{\bibfnamefont{V.}~\bibnamefont{Dupuis}},
  \bibinfo{author}{\bibfnamefont{E.}~\bibnamefont{Vincent}},
  \bibinfo{author}{\bibfnamefont{M.}~\bibnamefont{Alba}}, \bibnamefont{and}
  \bibinfo{author}{\bibfnamefont{J.}~\bibnamefont{Hammann}},
  \bibinfo{journal}{Eur. Phys. J. B} \textbf{\bibinfo{volume}{29}},
  \bibinfo{pages}{19} (\bibinfo{year}{2002}).

\bibitem[{\citenamefont{E.Vincent et~al.}(2000)\citenamefont{E.Vincent,
  V.Dupuis, M.Alba, J.Hammann, and
  J.-P.Bouchaud}}]{Vincent:(Hammann:Bouchaud):00}
\bibinfo{author}{\bibnamefont{E.Vincent}},
  \bibinfo{author}{\bibnamefont{V.Dupuis}},
  \bibinfo{author}{\bibnamefont{M.Alba}},
  \bibinfo{author}{\bibnamefont{J.Hammann}}, \bibnamefont{and}
  \bibinfo{author}{\bibnamefont{J.-P.Bouchaud}}, \bibinfo{journal}{Europhys.
  Lett.} \textbf{\bibinfo{volume}{50}}, \bibinfo{pages}{674}
  (\bibinfo{year}{2000}).

\bibitem[{\citenamefont{Colla et~al.}(2000)\citenamefont{Colla, Chao, Weissman,
  and Viehland}}]{Colla(Chao:mbw):00}
\bibinfo{author}{\bibfnamefont{E.~V.} \bibnamefont{Colla}},
  \bibinfo{author}{\bibfnamefont{L.~K.} \bibnamefont{Chao}},
  \bibinfo{author}{\bibfnamefont{M.~B.} \bibnamefont{Weissman}},
  \bibnamefont{and} \bibinfo{author}{\bibfnamefont{D.~D.}
  \bibnamefont{Viehland}}, \bibinfo{journal}{Phys. Rev. Lett.}
  \textbf{\bibinfo{volume}{85}}, \bibinfo{pages}{3033} (\bibinfo{year}{2000}).

\bibitem[{\citenamefont{Bellon et~al.}(2000)\citenamefont{Bellon, Ciliberto,
  and Laroche}}]{Bellon:00:EPL}
\bibinfo{author}{\bibfnamefont{L.}~\bibnamefont{Bellon}},
  \bibinfo{author}{\bibfnamefont{S.}~\bibnamefont{Ciliberto}},
  \bibnamefont{and} \bibinfo{author}{\bibfnamefont{C.}~\bibnamefont{Laroche}},
  \bibinfo{journal}{Europhys. Lett.} \textbf{\bibinfo{volume}{51}},
  \bibinfo{pages}{551} (\bibinfo{year}{2000}).

\bibitem[{\citenamefont{Mueller and Shchur}(2004)}]{Mueller:04:EPL}
\bibinfo{author}{\bibfnamefont{V.}~\bibnamefont{Mueller}} \bibnamefont{and}
  \bibinfo{author}{\bibfnamefont{Y.}~\bibnamefont{Shchur}},
  \bibinfo{journal}{Europhys. Lett.} \textbf{\bibinfo{volume}{65}},
  \bibinfo{pages}{137} (\bibinfo{year}{2004}).

\bibitem[{\citenamefont{Alberici-Kious
  et~al.}(1998)\citenamefont{Alberici-Kious, Bouchaud, Cugliandolo, Doussineau,
  and Levelut}}]{Alberici-Kious:98:PRL}
\bibinfo{author}{\bibfnamefont{F.}~\bibnamefont{Alberici-Kious}},
  \bibinfo{author}{\bibfnamefont{J.~P.} \bibnamefont{Bouchaud}},
  \bibinfo{author}{\bibfnamefont{L.~F.} \bibnamefont{Cugliandolo}},
  \bibinfo{author}{\bibfnamefont{P.}~\bibnamefont{Doussineau}},
  \bibnamefont{and} \bibinfo{author}{\bibfnamefont{A.}~\bibnamefont{Levelut}},
  \bibinfo{journal}{Phys. Rev. Lett.} \textbf{\bibinfo{volume}{81}},
  \bibinfo{pages}{4987} (\bibinfo{year}{1998}).

\bibitem[{\citenamefont{Bouchaud et~al.}(2001)\citenamefont{Bouchaud,
  Doussineau, de~Lacerda-Aroso, and Levelut}}]{Bouchaud:EPJB:01}
\bibinfo{author}{\bibfnamefont{J.~P.} \bibnamefont{Bouchaud}},
  \bibinfo{author}{\bibfnamefont{P.}~\bibnamefont{Doussineau}},
  \bibinfo{author}{\bibfnamefont{T.}~\bibnamefont{de~Lacerda-Aroso}},
  \bibnamefont{and} \bibinfo{author}{\bibfnamefont{A.}~\bibnamefont{Levelut}},
  \bibinfo{journal}{Eur. Phys. J. B} \textbf{\bibinfo{volume}{21}},
  \bibinfo{pages}{335} (\bibinfo{year}{2001}).

\bibitem[{\citenamefont{Vakhrushev et~al.}(1996)\citenamefont{Vakhrushev,
  Nabereznov, Sinha, Feng, and Egami}}]{Vakhrushev(Egami):96}
\bibinfo{author}{\bibfnamefont{S.}~\bibnamefont{Vakhrushev}},
  \bibinfo{author}{\bibfnamefont{A.}~\bibnamefont{Nabereznov}},
  \bibinfo{author}{\bibfnamefont{S.~K.} \bibnamefont{Sinha}},
  \bibinfo{author}{\bibfnamefont{Y.~P.} \bibnamefont{Feng}}, \bibnamefont{and}
  \bibinfo{author}{\bibfnamefont{T.}~\bibnamefont{Egami}}, \bibinfo{journal}{J.
  Phys. Chem. Solids} \textbf{\bibinfo{volume}{57}}, \bibinfo{pages}{1517}
  (\bibinfo{year}{1996}).

\bibitem[{\citenamefont{Colla et~al.}(2002)\citenamefont{Colla, Chao, and
  Weissman}}]{Colla(Chao:mbw):02}
\bibinfo{author}{\bibfnamefont{E.~V.} \bibnamefont{Colla}},
  \bibinfo{author}{\bibfnamefont{L.~K.} \bibnamefont{Chao}}, \bibnamefont{and}
  \bibinfo{author}{\bibfnamefont{M.~B.} \bibnamefont{Weissman}},
  \bibinfo{journal}{Phys. Rev. Lett.} \textbf{\bibinfo{volume}{88}},
  \bibinfo{pages}{017601} (\bibinfo{year}{2002}).

\bibitem[{\citenamefont{Weissman et~al.}(2003)\citenamefont{Weissman, Colla,
  and Chao}}]{Weissman:03Williamsburg}
\bibinfo{author}{\bibfnamefont{M.~B.} \bibnamefont{Weissman}},
  \bibinfo{author}{\bibfnamefont{E.~V.} \bibnamefont{Colla}}, \bibnamefont{and}
  \bibinfo{author}{\bibfnamefont{L.~K.} \bibnamefont{Chao}}, in
  \emph{\bibinfo{booktitle}{Fundamental Physics of Ferroelectrics 2003}},
  edited by \bibinfo{editor}{\bibfnamefont{P.~K.} \bibnamefont{Davies}}
  \bibnamefont{and} \bibinfo{editor}{\bibfnamefont{D.~J.} \bibnamefont{Singh}}
  (\bibinfo{year}{2003}), vol. \bibinfo{volume}{677}, pp.
  \bibinfo{pages}{33--40}.

\bibitem[{\citenamefont{Colla and Weissman}(2005)}]{Colla(mbw):05}
\bibinfo{author}{\bibfnamefont{E.~V.} \bibnamefont{Colla}} \bibnamefont{and}
  \bibinfo{author}{\bibfnamefont{M.~B.} \bibnamefont{Weissman}},
  \bibinfo{journal}{Phys. Rev. B} \textbf{\bibinfo{volume}{72}},
  \bibinfo{pages}{104106} (\bibinfo{year}{2005}).

\bibitem[{\citenamefont{Ryan et~al.}(1987)\citenamefont{Ryan, Coey, Batalla,
  Altounian, and Strom-Olsen}}]{Ryan(Coey):87}
\bibinfo{author}{\bibfnamefont{D.~H.} \bibnamefont{Ryan}},
  \bibinfo{author}{\bibfnamefont{J.~M.~D.} \bibnamefont{Coey}},
  \bibinfo{author}{\bibfnamefont{E.}~\bibnamefont{Batalla}},
  \bibinfo{author}{\bibfnamefont{Z.}~\bibnamefont{Altounian}},
  \bibnamefont{and} \bibinfo{author}{\bibfnamefont{J.~O.}
  \bibnamefont{Strom-Olsen}}, \bibinfo{journal}{Phys. Rev. B}
  \textbf{\bibinfo{volume}{35}}, \bibinfo{pages}{8630} (\bibinfo{year}{1987}).

\bibitem[{\citenamefont{Senoussi et~al.}(1988)\citenamefont{Senoussi, Hadjoudj,
  Jouret, Bilotte, and Fourmeaux}}]{Senoussi:88}
\bibinfo{author}{\bibfnamefont{S.}~\bibnamefont{Senoussi}},
  \bibinfo{author}{\bibfnamefont{S.}~\bibnamefont{Hadjoudj}},
  \bibinfo{author}{\bibfnamefont{P.}~\bibnamefont{Jouret}},
  \bibinfo{author}{\bibfnamefont{J.}~\bibnamefont{Bilotte}}, \bibnamefont{and}
  \bibinfo{author}{\bibfnamefont{R.}~\bibnamefont{Fourmeaux}},
  \bibinfo{journal}{J. Appl. Phys.} \textbf{\bibinfo{volume}{63}},
  \bibinfo{pages}{4086} (\bibinfo{year}{1988}).

\bibitem[{\citenamefont{Dkhiil et~al.}(2001)\citenamefont{Dkhiil, Kiat,
  Calvarin, Baldinozzi, Vakhrushev, and Suard}}]{Dkhil:01}
\bibinfo{author}{\bibfnamefont{B.}~\bibnamefont{Dkhiil}},
  \bibinfo{author}{\bibfnamefont{J.~M.} \bibnamefont{Kiat}},
  \bibinfo{author}{\bibfnamefont{G.}~\bibnamefont{Calvarin}},
  \bibinfo{author}{\bibfnamefont{G.}~\bibnamefont{Baldinozzi}},
  \bibinfo{author}{\bibfnamefont{S.~B.} \bibnamefont{Vakhrushev}},
  \bibnamefont{and} \bibinfo{author}{\bibfnamefont{E.}~\bibnamefont{Suard}},
  \bibinfo{journal}{Phys. Rev. B} \textbf{\bibinfo{volume}{65}},
  \bibinfo{pages}{024104:1} (\bibinfo{year}{2001}).

\bibitem[{\citenamefont{Egami}(1999)}]{Egami:99}
\bibinfo{author}{\bibfnamefont{T.}~\bibnamefont{Egami}},
  \bibinfo{journal}{Ferroelectrics} \textbf{\bibinfo{volume}{222}},
  \bibinfo{pages}{163} (\bibinfo{year}{1999}).

\bibitem[{\citenamefont{Vakhrushev et~al.}(1997)\citenamefont{Vakhrushev, Kiat,
  and Dkhil}}]{Vakhrushev:97}
\bibinfo{author}{\bibfnamefont{S.~B.} \bibnamefont{Vakhrushev}},
  \bibinfo{author}{\bibfnamefont{J.-M.} \bibnamefont{Kiat}}, \bibnamefont{and}
  \bibinfo{author}{\bibfnamefont{B.}~\bibnamefont{Dkhil}},
  \bibinfo{journal}{Solid State Commun.} \textbf{\bibinfo{volume}{103}},
  \bibinfo{pages}{477} (\bibinfo{year}{1997}).

\bibitem[{\citenamefont{Blinc et~al.}(2003)\citenamefont{Blinc, Laguta, and
  Zalar}}]{Blinc:03}
\bibinfo{author}{\bibfnamefont{R.}~\bibnamefont{Blinc}},
  \bibinfo{author}{\bibfnamefont{V.}~\bibnamefont{Laguta}}, \bibnamefont{and}
  \bibinfo{author}{\bibfnamefont{B.}~\bibnamefont{Zalar}},
  \bibinfo{journal}{Phys. Rev. Lett.} \textbf{\bibinfo{volume}{91}},
  \bibinfo{pages}{247601} (\bibinfo{year}{2003}).

\bibitem[{\citenamefont{Jeong et~al.}(2005)\citenamefont{Jeong, Darling, Lee,
  Proffen, Heffner, Park, Hong, Dmowski, and Egami}}]{Jeong:05:PRL}
\bibinfo{author}{\bibfnamefont{I.~K.} \bibnamefont{Jeong}},
  \bibinfo{author}{\bibfnamefont{T.~W.} \bibnamefont{Darling}},
  \bibinfo{author}{\bibfnamefont{J.~K.} \bibnamefont{Lee}},
  \bibinfo{author}{\bibfnamefont{T.}~\bibnamefont{Proffen}},
  \bibinfo{author}{\bibfnamefont{R.~H.} \bibnamefont{Heffner}},
  \bibinfo{author}{\bibfnamefont{J.~S.} \bibnamefont{Park}},
  \bibinfo{author}{\bibfnamefont{K.~S.} \bibnamefont{Hong}},
  \bibinfo{author}{\bibfnamefont{W.}~\bibnamefont{Dmowski}}, \bibnamefont{and}
  \bibinfo{author}{\bibfnamefont{T.}~\bibnamefont{Egami}},
  \bibinfo{journal}{Phys. Rev. Lett.} \textbf{\bibinfo{volume}{94}},
  \bibinfo{pages}{147602} (\bibinfo{year}{2005}).

\bibitem[{\citenamefont{Viehland et~al.}(1995)\citenamefont{Viehland, Kim, and
  Li}}]{Viehland:95:APL}
\bibinfo{author}{\bibfnamefont{D.}~\bibnamefont{Viehland}},
  \bibinfo{author}{\bibfnamefont{M.-C.} \bibnamefont{Kim}}, \bibnamefont{and}
  \bibinfo{author}{\bibfnamefont{J.-F.} \bibnamefont{Li}},
  \bibinfo{journal}{Appl. Phys. Lett.} \textbf{\bibinfo{volume}{67}},
  \bibinfo{pages}{2471} (\bibinfo{year}{1995}).

\bibitem[{\citenamefont{Xunhu et~al.}(1994)\citenamefont{Xunhu, Xu, and
  Viehland}}]{Dai:94:JAP}
\bibinfo{author}{\bibfnamefont{D.}~\bibnamefont{Xunhu}},
  \bibinfo{author}{\bibfnamefont{Z.}~\bibnamefont{Xu}}, \bibnamefont{and}
  \bibinfo{author}{\bibfnamefont{D.}~\bibnamefont{Viehland}},
  \bibinfo{journal}{Philos. Mag. B} \textbf{\bibinfo{volume}{70}},
  \bibinfo{pages}{33} (\bibinfo{year}{1994}).

\bibitem[{\citenamefont{Kartha et~al.}(1991)\citenamefont{Kartha, Castan,
  Krumhansl, and Sethna}}]{Kartha:91:PRL}
\bibinfo{author}{\bibfnamefont{S.}~\bibnamefont{Kartha}},
  \bibinfo{author}{\bibfnamefont{T.}~\bibnamefont{Castan}},
  \bibinfo{author}{\bibfnamefont{J.~A.} \bibnamefont{Krumhansl}},
  \bibnamefont{and} \bibinfo{author}{\bibfnamefont{J.~P.}
  \bibnamefont{Sethna}}, \bibinfo{journal}{Phys. Rev. Lett.}
  \textbf{\bibinfo{volume}{67}}, \bibinfo{pages}{3630} (\bibinfo{year}{1991}).

\bibitem[{\citenamefont{Sethna et~al.}(1992)\citenamefont{Sethna, Kartha,
  Castan, and Krumhansl}}]{Sethna:92:PhysScript}
\bibinfo{author}{\bibfnamefont{J.~P.} \bibnamefont{Sethna}},
  \bibinfo{author}{\bibfnamefont{S.}~\bibnamefont{Kartha}},
  \bibinfo{author}{\bibfnamefont{T.}~\bibnamefont{Castan}}, \bibnamefont{and}
  \bibinfo{author}{\bibfnamefont{J.~A.} \bibnamefont{Krumhansl}},
  \bibinfo{journal}{Phys. Scr.} \textbf{\bibinfo{volume}{T42}},
  \bibinfo{pages}{214} (\bibinfo{year}{1992}).

\bibitem[{\citenamefont{Balents et~al.}(1996)\citenamefont{Balents, Bouchaud,
  and Mezard}}]{Balents:96:JPhysI}
\bibinfo{author}{\bibfnamefont{L.}~\bibnamefont{Balents}},
  \bibinfo{author}{\bibfnamefont{J.-P.} \bibnamefont{Bouchaud}},
  \bibnamefont{and} \bibinfo{author}{\bibfnamefont{M.}~\bibnamefont{Mezard}},
  \bibinfo{journal}{J. Phys. I France} \textbf{\bibinfo{volume}{6}},
  \bibinfo{pages}{1007} (\bibinfo{year}{1996}).

\bibitem[{\citenamefont{Chao et~al.}(2003)\citenamefont{Chao, Colla, and
  Weissman}}]{Chao(mbw):03}
\bibinfo{author}{\bibfnamefont{L.~K.} \bibnamefont{Chao}},
  \bibinfo{author}{\bibfnamefont{E.~V.} \bibnamefont{Colla}}, \bibnamefont{and}
  \bibinfo{author}{\bibfnamefont{M.~B.} \bibnamefont{Weissman}}, in
  \emph{\bibinfo{booktitle}{Proceedings of SPIE}}, edited by
  \bibinfo{editor}{\bibfnamefont{M.~B.} \bibnamefont{Weissman}},
  \bibinfo{editor}{\bibfnamefont{N.~E.} \bibnamefont{Israeloff}},
  \bibnamefont{and} \bibinfo{editor}{\bibfnamefont{A.~S.} \bibnamefont{Kogan}}
  (\bibinfo{year}{2003}), vol. \bibinfo{volume}{5112}, pp.
  \bibinfo{pages}{325--330}.

\bibitem[{\citenamefont{Viehland et~al.}(2004)\citenamefont{Viehland, Li, and
  Colla}}]{Viehland:04:JAP}
\bibinfo{author}{\bibfnamefont{D.}~\bibnamefont{Viehland}},
  \bibinfo{author}{\bibfnamefont{J.}~\bibnamefont{Li}}, \bibnamefont{and}
  \bibinfo{author}{\bibfnamefont{E.~V.} \bibnamefont{Colla}},
  \bibinfo{journal}{J. Appl. Phys.} \textbf{\bibinfo{volume}{96}},
  \bibinfo{pages}{3379} (\bibinfo{year}{2004}).

\end{thebibliography}



\end{document}